\documentclass[12pt]{article}
\pdfoutput=1
\usepackage{graphics}
\usepackage{graphicx}
\usepackage{natbib}
\usepackage{bm}
\usepackage[letterpaper]{geometry}
\usepackage{amsmath}
\usepackage{algpseudocode}

\algblock{ParFor}{EndParFor}
\algnewcommand\algorithmicparfor{\textbf{parallel for}}
\algnewcommand\algorithmicpardo{\textbf{do}}
\algnewcommand\algorithmicendparfor{\textbf{end\ parallel for}}
\algrenewtext{ParFor}[1]{\algorithmicparfor\ #1\ \algorithmicpardo}
\algrenewtext{EndParFor}{\algorithmicendparfor}

\begin{document}
\title{Parallelizing Gaussian Process Calculations\\in R}

\author{Christopher J. Paciorek, Benjamin Lipshitz, Wei Zhuo, \\ Prabhat, Cari G. Kaufman, Rollin C. Thomas}
\maketitle 
\begin{centering}
\small{
Department of Statistics, University of California, Berkeley,\\
Department of Electrical Engineering and Computer Science, University of California, Berkeley,\\
College of Computing, Georgia Institute of Technology, \\
Computational Research Division, Lawrence Berkeley National Laboratory,\\
Department of Statistics, University of California, Berkeley, \\
Computational Cosmology Center, Lawrence Berkeley National Laboratory\\}
\end{centering}

\begin{abstract}
  We consider parallel computation for Gaussian process calculations
  to overcome computational and memory constraints 
  on the size of datasets that can be analyzed.  Using a hybrid parallelization approach that uses both
  threading (shared memory) and message-passing (distributed memory), we
  implement the core linear algebra operations used in spatial
  statistics and Gaussian process regression in an \texttt{R} package called \texttt{bigGP} that relies on \texttt{C} and \texttt{MPI}. The approach divides the matrix into blocks such that the computational
load is balanced across processes while communication between processes is limited. The package
  provides an API enabling \texttt{R} programmers to implement Gaussian process-based
  methods by using the distributed linear algebra operations without any \texttt{C} or \texttt{MPI} coding.
  We illustrate the approach and software by analyzing an astrophysics
  dataset with $n=67,275$ observations.
\end{abstract}


\section{Introduction}
Gaussian processes are widely used in statistics and machine learning
 for spatial and spatio-temporal modeling \citep{Bane:etal:2003}, design and analysis of computer experiments \citep{Kenn:OHag:2001},
and non-parametric regression \citep{Rasm:Will:2006}. 
One popular example is the spatial statistics method of kriging, which is equivalent to conditional expectation under a Gaussian process model for the  unknown spatial field.
However standard implementations of Gaussian process-based methods 
are computationally intensive because they involve calculations with
covariance matrices of size $n$ by $n$ where $n$ is the number of
locations with observations. In particular the computational
bottleneck is generally the Cholesky decomposition of the covariance
matrix, whose computational cost is of order $n^3$. 

For example, a
basic spatial statistics model (in particular a geostatistical model) can be specified in a hierarchical fashion as
\begin{eqnarray*}
\bm{Y}|\bm{g},\bm{\theta} & \sim & \mathcal{N}(\bm{g},\bm{C}_y(\bm{\theta}))\\
\bm{g}|\bm{\theta} & \sim & \mathcal{N}(\bm{\mu}(\bm{\theta}),\bm{C}_g(\bm{\theta}))\end{eqnarray*}
 where $g$ is a vector of latent spatial process values at the $n$ locations, $\bm{C}_y(\bm{\theta})$ is an error covariance matrix (often diagonal), $\mu(\bm{\theta})$ is the mean vector of the latent process, $\bm{C}_g(\bm{\theta})$ is the spatial covariance matrix of the latent process, and $\bm{\theta}$ is a vector of unknown parameters.
We can marginalize over $g$ to obtain the marginal likelihood
\[
\bm{Y}|\bm{\theta} \sim\mbox{N}(\bm{\mu}(\bm{\theta}), \bm{C}(\bm{\theta}))\]
where $\bm{C}(\bm{\theta}) = \bm{C}_y(\bm{\theta}) + \bm{C}_g(\bm{\theta})$. 
This gives us the marginal density,
\[f(\bm{y})\propto|\bm{C}(\bm{\theta})|^{-1/2}\exp\left\{
  -\frac{1}{2}(\bm{y}-\bm{\mu}(\bm{\theta}))^{\top}(\bm{C}(\bm{\theta}))^{-1}(\bm{y}-\bm{\mu}(\bm{\theta}))\right\}, \]
which is maximized over $\bm{\theta}$ t ofind the maximum
likelihood estimator. At each iteration 
in maximization of the log-likelihood, the expensive computations are to compute the entries of the matrix $\bm{C}(\bm{\theta})$ as a function of $\bm{\theta}$, calculate the Cholesky decomposition,
$\bm{LL}^{T}=\bm{C}(\bm{\theta})$, and solve a system of
equations $\bm{L}^{-1}(\bm{y}-\bm{\mu}(\bm{\theta}))$ via a backsolve operation. Given the MLE, $\hat{\bm{\theta}}$,
one might then do spatial prediction, calculate the variance of the prediction, and simulate
realizations conditional on the data. These additional tasks involve the same expensive computations plus
a few additional closely-related computations.  

In general the Cholesky decomposition will be the rate-limiting step
in these tasks, although calculation of covariance matrices can
also be a bottleneck.  In addition to computational limitations,
memory use can be a limitation, as storage of the covariance matrix
involves $n^2$ floating points. For example, simply storing a covariance matrix for $n=20,000$
observations in memory uses approximately 3.2 GB of RAM. As a result of the computational and memory limitations, standard
spatial statistics methods are typically applied to datasets with at most a few thousand observations.

To overcome these limitations, a
small industry has arisen to develop computationally-efficient approaches to spatial
statistics, involving reduced rank approximations
\citep{Kamm:Wand:2003,Bane:etal:2008,Cres:Joha:2008}, tapering the
covariance matrix to induce sparsity
\citep{Furr:etal:2006,Kauf:etal:2008}, approximation of the likelihood
\citep{Stei:etal:2004}, and fitting local models by stratifying the spatial
domain \citep{Gram:etal:2008}, among others. At the same time, computer scientists
have developed and implemented parallel linear algebra algorithms that use modern distributed memory and multi-core hardware.
Rather than modifying the statistical
model, as statisticians have focused on, here we consider the use of
parallel algorithms to overcome computational limitations, enabling
analyses with much larger covariance matrices than would be otherwise possible.

We present an algorithm and \texttt{R} package, \texttt{bigGP}, for distributed linear algebra
calculations focused on those used in spatial statistics and
closely-related Gaussian process regression methods. The approach
divides the covariance matrix (and other necessary matrices and
vectors) into blocks, with the blocks distributed amongst processors
in a distributed computing environment.  The algorithm builds on that
encoded within the widely-used parallel linear algebra package,
\texttt{ScaLAPACK} \citep{scalapack}, a parallel extension to the standard LAPACK \citep{lapack} routines. The
core functions in the \texttt{bigGP} package are \texttt{C} functions, with \texttt{R} wrappers, that rely on standard
BLAS \citep{blas} functionality and on \texttt{MPI} for message passing.  This set of core
functions includes Cholesky decomposition, forward and backsolve, and crossproduct calculations. These functions, plus some auxiliary functions for communication of inputs
and outputs to the processes, provide an API through which an \texttt{R}
programmer can implement methods for Gaussian-process-based
computations. Using the API, we provide a set of methods for the
standard operations involved in kriging and Gaussian process
regression, namely 
\begin{itemize}
\item likelihood optimization, 
\item prediction,
\item calculation of prediction uncertainty, 
\item unconditional simulation of
Gaussian processes, and 
\item conditional simulation given data.
\end{itemize}
These
methods are provided as \texttt{R} functions in the package. We
illustrate the use of the software for Gaussian process regression in
an astrophysics application.


\section{Parallel algorithm and software implementation}

\subsection{Distributed linear algebra calculations}
Parallel computation can be done in both shared memory and distributed memory contexts. Each uses multiple CPUs. In a shared memory context (such as computers with one or more chips with multiple cores), multiple CPUs have access to the same memory and so-called 'threaded' calculations can be done, in which code is written (e.g., using the openMP protocol) to use more than one CPU at once to carry out a task, with each CPU having access to the objects in memory. In a distributed memory context, one has a collection of nodes, each with their own memory. Any information that must be shared with other nodes must be done via message-passing, such as using the \texttt{MPI} standard. Our distributed calculations use both threading and message-passing to exploit the capabilities of modern computing clusters with multiple-core nodes.

\begin{figure}
 \centering
\includegraphics[scale=.6]{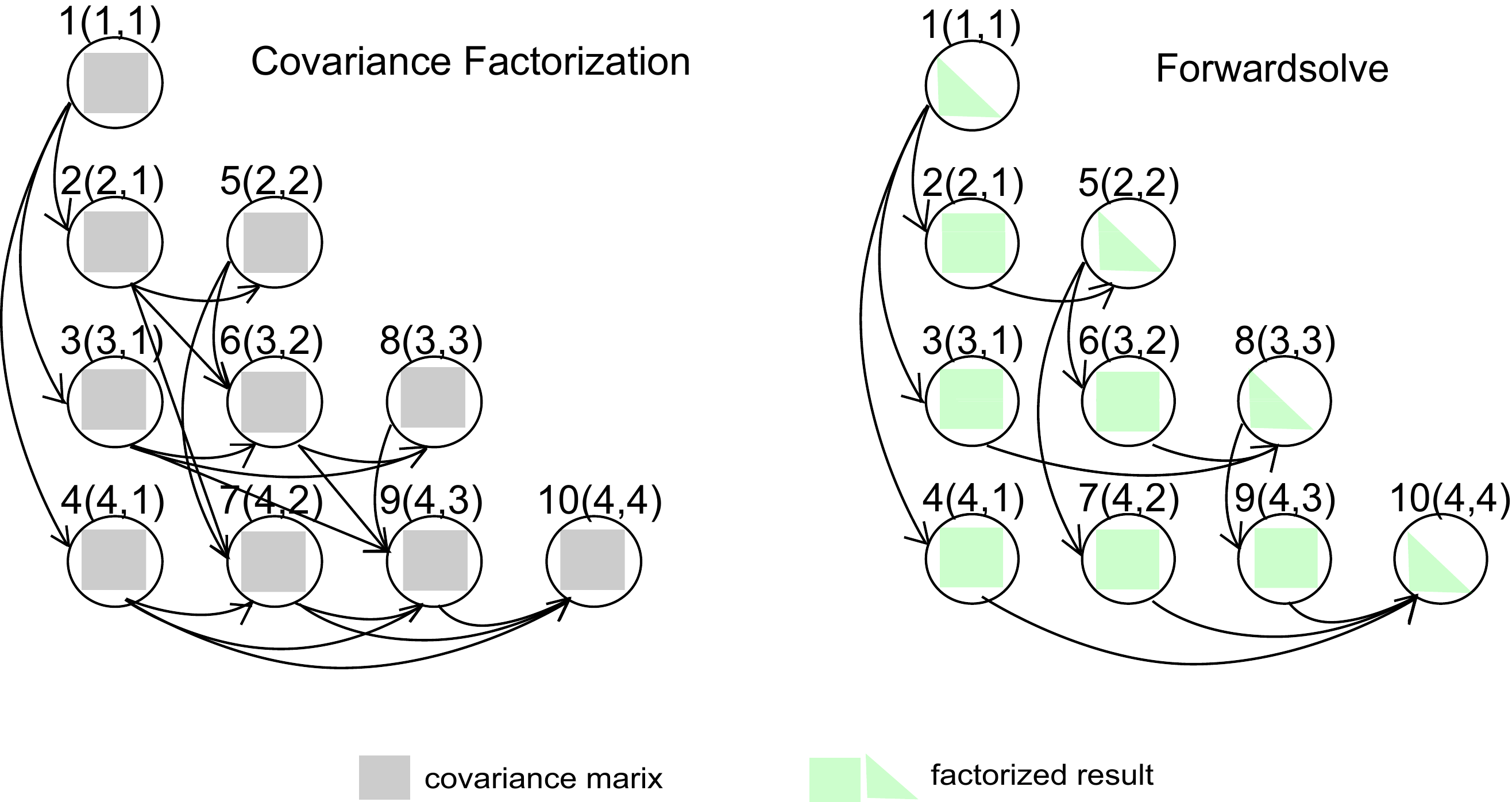}
 \caption{Dependency graphs for the distributed Cholesky factorization and forwardsolve. The labels of the form ``X(Y,Z)'' indicate the process ID (X) and Cartesian coordinate identifier of the process (Y,Z).\label{fig:dependency-graph}}
\end{figure}

We begin by describing a basic parallel Cholesky decomposition, which is done on blocks of the matrix and is implemented in \texttt{ScaLAPACK}. Fig.~\ref{fig:dependency-graph} shows a schematic of the block-wise Cholesky factorization, as well as the forwardsolve operation, where a triangular matrix is divided into 10 blocks, a $B=4$ by $B=4$ array of blocks. The arrows show the dependence of each block on the other blocks; an arrow connecting two blocks stored on different nodes indicates that communication is necessary between those nodes.  For the Cholesky decomposition, the calculations for the diagonal blocks all involve Cholesky decompositions (and symmetric matrix multiplication for all but the first process), while those for the off-diagonal blocks involve forwardsolve operations and (for all but the first column of processes) matrix multiplication and subtraction.  Most of the total work is in the matrix multiplications.

The first block must be factored before the forwardsolve can be applied to blocks 2-4. After the forwardsolves, all the remaining blocks can be updated with a matrix multiplication and subtraction.  At this point the decomposition of blocks 1-4 is complete and they will not be used for the rest of the computation.  This process is repeated along each block column, so for example blocks 8-10 must wait for blocks 6 and 7 to finish before all the necessary components are available to finish their own calculations.

To specify the distributed algorithm, there are several choices to be made: the number of blocks, $B$, how to distribute these blocks among the processes, how to distribute these processes among the nodes, and how many nodes to use.  We discuss the tradeoffs involved in these choices below and the choices that our algorithm makes in Section~\ref{sec:our-alg}.

Given a matrix of a given size, specifying the number of blocks is equivalent to choosing the size of the blocks.  Larger blocks allow for more efficient local computations and less total communication.  The largest effect here is the on-node cache subsystem, which allows each node to run near its peak performance only if the ratio of computation to memory traffic is high enough.  The computational efficiency will increase with the block size until the blocks are large enough to fill the cache available to one process.  For example, if 8MB of cache are available, one would like to have block size at least $1024\times1024$.  However, smaller blocks allow the algorithm to better balance the computational load between the processes (and therefore ultimately among the cores and nodes of the computer) by assigning multiple blocks to each process.  The first block must finish before anything else can be done; assuming that each block is assigned to only one process, all the other processes must wait for this block to be factored before they can begin computation.  More generally, the diagonal and first off-diagonal blocks form a critical path of the algorithm.  In Fig.~\ref{fig:dependency-graph}, this critical path is of blocks 1, 2, 5, 6, 8, 9, 10.  The decomposition, forwardsolves, and multiplicative updates of each of these blocks must be done sequentially.  Decreasing the block size decreases the amount of work along this critical path, thus improving the load balance.  Put another way, decreasing the block size decreases how long the majority of the processes wait for the processes in a given column to finish before they can use the results from that column to perform their own computation.

Given a matrix of a fixed size and a fixed number of nodes, if we were to use the maximum block size, we would distribute one block per process and one process per node.  If we use a smaller block size, we can accommodate the extra blocks either by assigning multiple blocks to a each process, or multiple processes to each node.  Consider first running just one process per node.  For the linear algebra computations, by using a threaded BLAS library (such as \texttt{openBLAS}, \texttt{MKL}, or \texttt{ACML}), it is still possible to attain good multi-core performance on a single process.  However, any calculations that are not threaded will not be able to use all of the cores on a given node, reducing computational efficiency. An example of this occurs in our \texttt{R} package where the user-defined covariance function (which is an \texttt{R} function) will typically not be threaded unless the user codes it in threaded \texttt{C} code (e.g., using openMP) and calls the \texttt{C} code from the \texttt{R} function or uses a parallel framework in \texttt{R}.

Alternatively, one could specify the block size and the number of nodes such that more than one process runs on each node. For non-threaded calculations, this manually divides the computation amongst multiple cores on a node, increasing efficiency. However it reduces the number of cores available for a given threaded calculation (presuming that cores are assigned exclusively to a single process, as when using the \texttt{openMPI} implementation with \texttt{Rmpi}) and may decrease efficiency by dividing calculations into smaller blocks with more message passing.  This is generally a satisfactory solution in a small cluster, but will lose efficiency past a few tens of nodes. Finally, one can assign multiple blocks per process, our chosen approach, described next.


\subsection{Our algorithm}
\label{sec:our-alg}

Our approach assigns one process per node, but each process is assigned multiple blocks of the matrices. This allows each process access to all the cores on a node to maximally exploit threading. We carefully choose which blocks are assigned to each process to achieve better load-balancing and limit communication.  We choose an efficient order for each process to carry out the operations for the blocks assigned to it.  Our package is flexible enough to allow the user to instead run multiple processes per node, which may improve the efficiency of the user-defined covariance function at the expense of higher communication costs in the linear algebra computations.

\begin{figure}
  \centering
  \includegraphics[scale=.75]{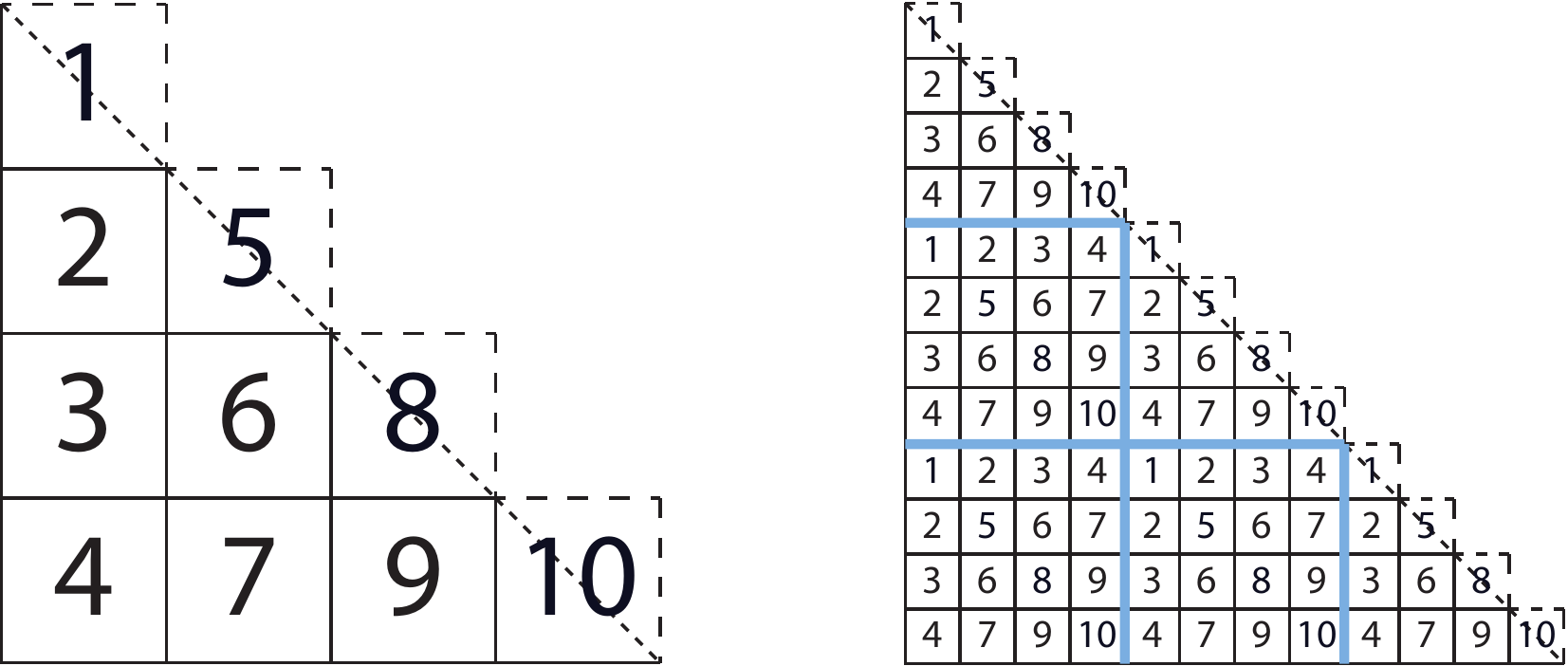}
  \caption{The matrix layout used by our algorithm with $D=4$ and $h=1$ (left) or $h=3$ (right).  The numbers indicate which process owns a given block.  When $h=1$, each of the 10 processes owns one block of the matrix.  When $h=3$, the blocks are a third the size in each dimension.  The diagonal processes (1, 5, 8, 10) each own $h(h+1)/2=6$ blocks, and the off-diagonal processes (2, 3, 4, 6, 7, 9) each own $h^2=9$ blocks.}
  \label{fig:matrix-vary-h}
\end{figure}

We require that the number of processes is $P=D(D+1)/2 \in \{1,3,6,10,15,\ldots\}$ for some integer value of $D$.  We introduce another quantity $h$ that determines how many blocks each process owns.  The number of blocks is given by $B=hD$, and so the block size is
$\left\lceil\frac{n}{hD}\right\rceil,$
where $n$ is the order of the matrix.  See Fig.~\ref{fig:matrix-vary-h} for an example of the layout with $D=4$ and either $h=1$ or $h=3$.  Each ``diagonal process'' has $h(h+1)/2$ blocks, and each ``off-diagonal process'' has $h^2$ blocks of the triangular matrix.

As discussed above, small values of $h$ increase on-node efficiency and reduce communication, but large values of $h$ improve load balance.  On current architectures, a good heuristic is to choose $h$ so that the block size is about $1000$, but the user is encouraged to experiment and determine what value works best for a given computer and problem size.  When using more than 8 cores per process, the block size should probably be increased.  Note that we pad the input matrix so that the number of rows and columns is a multiple of $hD$.  This padding will have a minimal effect on the computation time; if the block size is chosen to be near 1000 as we suggest, the padding will be at most one part in a thousand.

Note that when $h>1$, there are essentially two levels of blocking, indicated by the thin black lines and the thick blue lines in Fig.~\ref{fig:matrix-vary-h}.  Our algorithm is guided by these blocks.  At a high level, the algorithm sequentially follows the Cholesky decomposition of the large (blue) blocks as described in the previous section.  Each large block is divided among all the processors, and all the processors participate in each step.  For example, the first step is to perform Cholesky decomposition on the first large block.  To do so, we follow exactly the $h=1$ algorithm (making use of the Cartesian coordinate identification system indicated in Fig.~\ref{fig:dependency-graph}):
\begin{algorithmic}[1]
\For{$i=1$ to $D$}
\State Processor $(i,i)$ computes the Cholesky decomposition of its block
\ParFor{$j=i+1$ to $D$}
\State Processor $(i,i)$ sends its block to processor $(j,i)$
\State Processor $(j,i)$ updates its block with a triangular solve
 \ParFor{$k=i+1$ to $D$}
 \If{$k\leq j$}
   \State Processor $(j,i)$ sends its block to processor $(j,k)$
 \Else
   \State Processor $(j,i)$ sends its block to processor $(k,j)$
 \EndIf
 \EndParFor
\EndParFor
\ParFor{$j=i+1$ to $D$}
 \ParFor{$k=j+1$ to $D$}
  \State Processor $(k,j)$ updates its block with a matrix multiplication
 \EndParFor
\EndParFor
\EndFor
\end{algorithmic}
The $h=1$ algorithm is poorly load balanced; for example going from $D=1$ to $D=2$ (one process to three processes), one would not expect any speedup because every operation is along the critical path.  However, because it is a small portion of the entire calculation for $h>1$, the effect on the total runtime is small.  Instead, most of the time is spent in matrix multiplications of the large blue blocks, which are well load-balanced.


\subsubsection{Memory use}
The number of entries in a triangular $n\times n$ matrix is $n(n+1)/2$.  Ideally, it would be possible to perform computations even if there is only barely this much memory available across all the nodes, that is if there were enough memory for $n(n+1)/(D(D+1))$ entries per node.  Our algorithm does not reach this ideal, but it has a small memory overhead that decreases as $D$ or $h$ increase. The maximum memory use is by the off-diagonal nodes that own $h^2$ blocks.  Additionally, during the course of the algorithm they must temporarily store up to 4 more blocks.  Assuming for simplicity that $hD$ evenly divides $n$, the maximum memory use on a node is then
\begin{align*}
M&\leq \left(\frac{n}{hD}\right)^2(h^2+4)=\frac{n(n+1)}{D(D+1)}\left(1+\frac{4nD+n^2h^2+4n-Dh^2}{Dh^2n+Dh^2}\right) \\&< \frac{n(n+1)}{D(D+1)}\left(1+\frac{4}{h^2}+\frac{1}{D}+\frac{4}{Dh^2}\right).
\end{align*}
For example when $h=3$ and $D=4$, the memory required is about 1.8 times the memory needed to hold a triangular matrix.  Increasing $h$ and $D$ decreases this overhead factor toward 1.


\subsubsection{Advantages of our approach}

So far we have focused our discussion on the Cholesky factorization, as this is generally the rate-limiting step in Gaussian process methods. Our approach and software also improve computational efficiency by exploiting the sequential nature of Gaussian process calculations, in which each task relies on a sequence of linear algebra calculations, many or all of which can be done in a distributed fashion. 
Our framework generates matrices in a distributed fashion, and keeps them distributed throughout a sequence of linear algebra computations, collecting results back to the master process only at the conclusion of the task. For example in likelihood calculation, we need not collect the full Cholesky factor at the master process but need only collect the (scalar) log-likelihood value that is computed using a sequence of distributed calculations (the Cholesky factorization, followed by a backsolve, calculation of a sum of squares, and calculation of a log-determinant). For prediction, if we have computed the Cholesky during likelihood maximization, we can use the distributed Cholesky as input to distributed forwardsolve and backsolve operations, collecting only the vector of predictions at the master process. This feature is critical both for avoiding the large communication overhead in collecting a matrix to a single processor and to allowing computations on matrices that are too big to fit on one node.


\texttt{ScaLAPACK} is an alternative to our approach and uses a very similar algorithmic approach. In Section \ref{sec:results} we show that our implementation is as fast or faster than using \texttt{ScaLAPACK} for the critical Cholesky decomposition.  In some ways our implemenatation is better optimized for triangular or symmetric matrices.  When storing symmetric matrices, \texttt{ScaLAPACK} requires memory space for the entire square matrix, whereas our implementation only requires a small amount more memory than the lower triangle takes.
Furthermore, unlike the \texttt{RScaLAPACK} interface (which is no longer available as a current \texttt{R} package) to \texttt{ScaLAPACK}, our implementation carries out multiple linear algebra calculations without collecting all the results back to the master process and calculates the covariance matrix in a distributed fashion.

Our software provides \texttt{R} users with a simple interface to distributed calculations that mimic the algorithmic implementation in \texttt{ScaLAPACK} while also providing \texttt{R} programmers with an API to enable access to the core distributed back-end linear algebra calculations, which are coded in \texttt{C} for efficiency.

\subsection{The bigGP R package}

\subsubsection{Overview}

The \texttt{R} package \texttt{bigGP} implements a set of core functions, all in a distributed fashion, that are useful for a variety of Gaussian process-based computational tasks. In particular we provide Cholesky factorization, forwardsolve, backsolve and multiplication operations, as well as a variety of auxiliary functions that are used with the core functions to implement high-level statistical tasks.  We also provide additional \texttt{R} functions for distributing objects to the processes, managing the objects, and collecting results at the master process.

This set of \texttt{R} functions provides an API for \texttt{R} developers. A developer can implement new tasks entirely in \texttt{R} without needing to know or use \texttt{C} or \texttt{MPI}. Indeed, using the API, we implement standard Gaussian process tasks: log-likelihood calculation, likelihood optimization, prediction, calculation of prediction uncertainty, unconditional simulation of Gaussian processes, and simulation of Gaussian process realizations conditional on data. Distributed cnstruction of mean vectors and covariance matrices is done using user-provided \texttt{R} functions that calculate the mean and covariance functions given a vector of parameters and arbitrary inputs. 

\subsubsection{API}

The API consists of 
\begin{itemize}
\item basic functions for listing and removing objects on the slave processes and copying objects to and from the slave processes: \texttt{remoteLs}, \texttt{remoteRm}, \texttt{push}, \texttt{pull},
\item functions for determining the lengths and indices of vectors and matrices assigned to a given slave process: \texttt{getDistributedVectorLength}, \\ \texttt{getDistributedTriangularMatrixLength}, \\ \texttt{getDistributedRectangularMatrixLength}, \texttt{remoteGetIndices},  
\item functions that distribute and collect objects to and from the slave processes, masking the details of how the objects are broken into pieces: \texttt{distributeVector}, \texttt{collectVector}, \texttt{collectDiagonal}, \\ \texttt{collectTriangularMatrix}, \texttt{collectRectangularMatrix}, and
\item functions that carry out linear algebra calculations on distributed vectors and matrices: \\ \texttt{remoteCalcChol}, \texttt{remoteForwardsolve}, \texttt{remoteBacksolve},\\ \texttt{remoteMultChol}, \texttt{remoteCrossProdMatVec}, \texttt{remoteCrossProdMatSelf}, \\ \texttt{remoteCrossProdMatSelfDiag}, \texttt{remoteConstructRnormVector}, \\ \texttt{remoteConstructRnormMatrix}. In addition there is a generic \texttt{remoteCalc} function that can carry out an arbitrary function call with either one or two inputs. 
\end{itemize}

The package must be initialized, which is done with the \texttt{bigGP.init} function. During initialization, slave processes are spawned and \texttt{R} packages loaded on the slaves, parallel random number generation is set up, and blocks are assigned to slaves, with this information stored on each slave process in the \texttt{.bigGP} object. Users need to start \texttt{R} in such a way (e.g., through a queueing system or via \texttt{mpirun}) that $P$ slave processes can be initialized, plus one for the master process, for a total of $P+1$. $P$ should be such that $P = D(D+1)/2$ for integer $D$, i.e., $P \in {3,6,10,15,\ldots}$. One may wish to have one process per node, with threaded calculations on each node via a threaded BLAS, or one process per core (in particular when a threaded BLAS is not available).

Our theoretical assessment and empirical tests suggest that the blocks of distributed matrices should be approximately of size $1000$ by $1000$. To achieve this, the package chooses $h$ given the number of observations, $n$, and the number of processes, $P$, such that the blocks are approximately that size, i.e., $n/(hD) \approx 1000$. However the user can override the default and we recommend that the user test different values on their system. 

\subsubsection{Kriging implementation}

The kriging implementation is built around two Reference Classes. 

The first is a \texttt{krigeProblem} class that contains metadata about the problem and manages the analysis steps. To set up the problem and distribute inputs to the processes, one instantiates an object in the class. The metadata includes the block replication factors and information about which calculations have been performed and which objects are  up-to-date (i.e., are consistent with the current parameter values). This allows the package to avoid repeating calculations when parameter values have not changed. Objects in the class are stored on the master process. 

The second is a \texttt{distributedKrigeProblem} class that contains the core distributed objects and information about which pieces of the distributed objects are stored in a given process. Objects in this class are stored on the slave processes. By using a ReferenceClass we create a namespace that avoids name conflicts amongst multiple problems, and we allow the distributed linear algebra functions to manipulate the (large) blocks by reference rather than by value.

The core methods of the \texttt{krigeProblem} class are a constructor; methods for constructing mean vectors and covariance matrices given user-provided mean and covariance functions; methods for calculating the log determinant, calculating the log density, optimizing the density with respect to the parameters, prediction (with prediction standard errors), finding the full prediction variance matrix, and simulating realizations conditional on the data. Note that from a Bayesian perspective, prediction is just calculation of the posterior mean, the prediction variance matrix is just the posterior variance, and simulation of realizations is just simulation from the posterior. All of these are conditional on the parameter estimates, so this can be viewed as empirical Bayes. 

It is possible to have multiple \texttt{krigeProblem} objects defined at once, with separate objects in memory and distributed amongst the processes. 
However, the partition factor, $D$, is constant within a given \texttt{R} session.

Code that uses the \texttt{krigeProblem} class to analyze an astrophysics data example is provided in Section~\ref{sec:example}.


\subsubsection{Using the API}

To extend the package to implement other Gaussian process methodologies, the two key elements are construction of the distributed objects and use of the core distributed linear algebra functions. Construction of the distributed objects should mimic the \\\texttt{localKrigeProblemConstructMean} and \texttt{localKrigeProblemConstructCov} functions in the package. These functions use user-provided functions that operate on a set of parameters, a list containing additional inputs, and a set of indices to construct the local piece of the object for the given indices. As a toy example, the package may set the indices of a matrix stored in the first process to be ${(1,1), (2,1), (1,2), (2,2)}$, namely the upper $2\times2$ block of a matrix. Given this set of indices, the user-provided function would need to compute these four elements of the matrix, which would then be stored as a vector, column-wise, in the process. Once the necessary vectors and matrices are computed, the distributed linear algebra functions allow one to manipulate the objects by name. As done with the \texttt{krigeProblem} class, we recommend the use of Reference Classes to store the various objects and functions associated with a given methodology. 

\section{Timing results}
\label{sec:results}

We focus on comparing computational speed for the Cholesky factorization, as this generally dominates the computational time for Gaussian process computations. We compare our implementation (in this case run as a distributed \texttt{C} program, as \texttt{R} serves only as a simple wrapper that calls the local Cholesky functions on the worker processes via the \texttt{mpi.remote.exec} function), for a variety of values of $h$, with \texttt{ScaLAPACK}. We consider several different dataset sizes and different numbers of nodes. We use Hopper, a Cray system hosted at the National Energy Research Scientific Computing center (NERSC). Each Hopper node consists of two 12-core AMD ``MagnyCours'' processors with 24 GB of memory.  Hopper jobs have access to a dedicated Cray Gemini interconnect to obtain low-latency and high bandwidth inter-process communication.  While Hopper has 24 cores per node, each node is divided into 4 NUMA regions each with 6 cores; in our experiments we try running one process per node, one process per NUMA region (4 per node), or one process per core (24 per node).

\subsection{Choice of h and comparison to ScaLAPACK}
\begin{figure}
  \centering
  \includegraphics[scale=1]{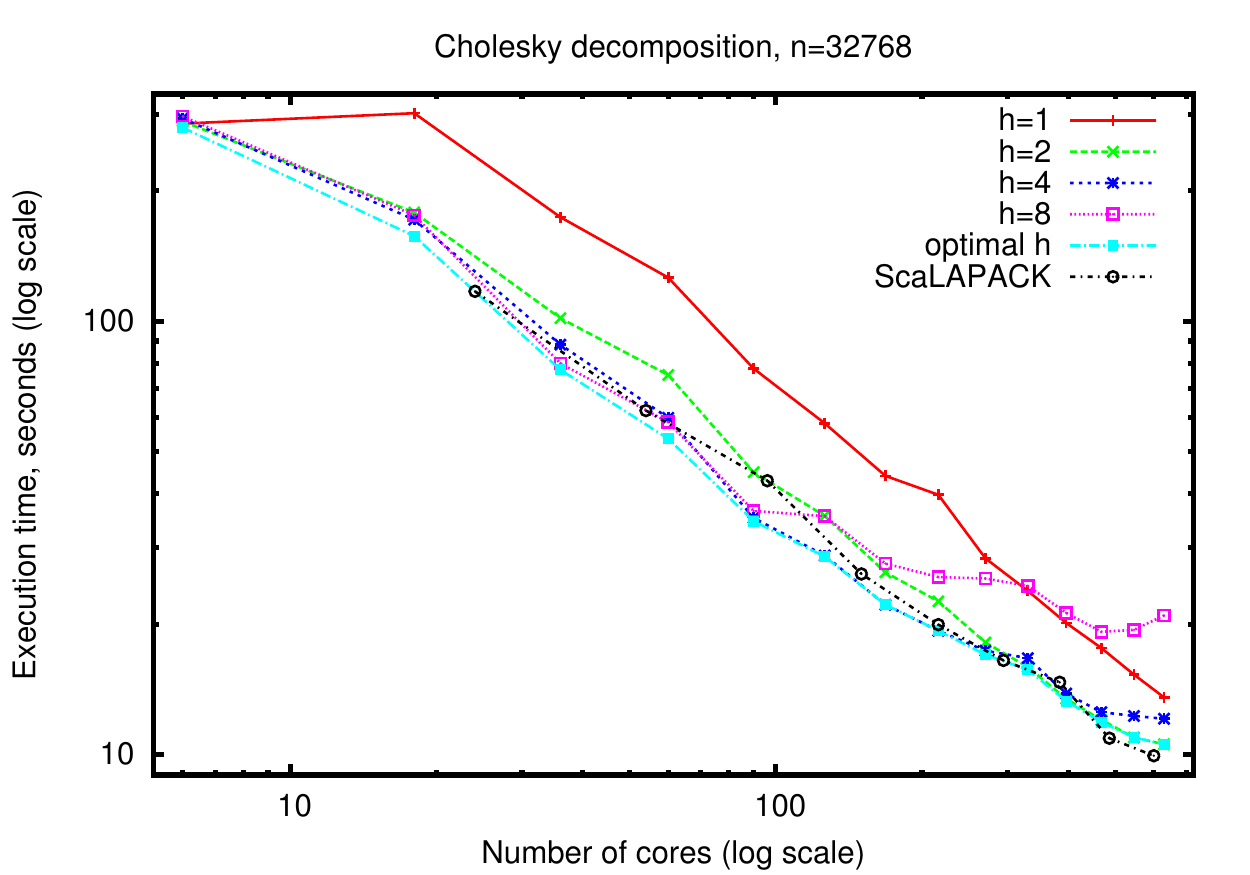}
  \caption{Runtimes for $32768\times32768$ Cholesky decomposition on Hopper with various values of $h$ using $6$ cores per process.  The last line shows \texttt{ScaLAPACK} as a benchmark.  The optimal value of $h$ was chosen by trying all values between 1 and 8.  The blocksize for \texttt{ScaLAPACK} corresponds to the best performance using a power of 2 blocks per process.}
  \label{fig:vary-h}
\end{figure}

In Fig.~\ref{fig:vary-h} we compare the performance at different values of $h$.  One notable feature is that for $h=1$ there is no performance improvement in increasing from $P=1$ to $P=3$, because there is no parallelism.  Allowing larger values of $h$ makes a speedup possible with $P=3$.  Generally, larger values of $h$ perform best when $P$ is small, but as $P$ grows the value of $h$ should decrease to keep the block size from getting too small.

Fig.~\ref{fig:vary-h} also compares our performance to \texttt{ScaLAPACK}, a standard distributed-memory linear algebra library.  Performance for \texttt{ScaLAPACK} (using the optimal block size) and our algorithm (using the optimal value of $h$) is similar.  We are thus able to get essentially the same performance on distributed linear algebra computations issued from \texttt{R} with our framework as if the programmer were working in \texttt{C} and calling \texttt{ScaLAPACK}.

\subsection{Timing with increasing problem size}
\begin{figure}
  \centering
  \includegraphics[scale=1]{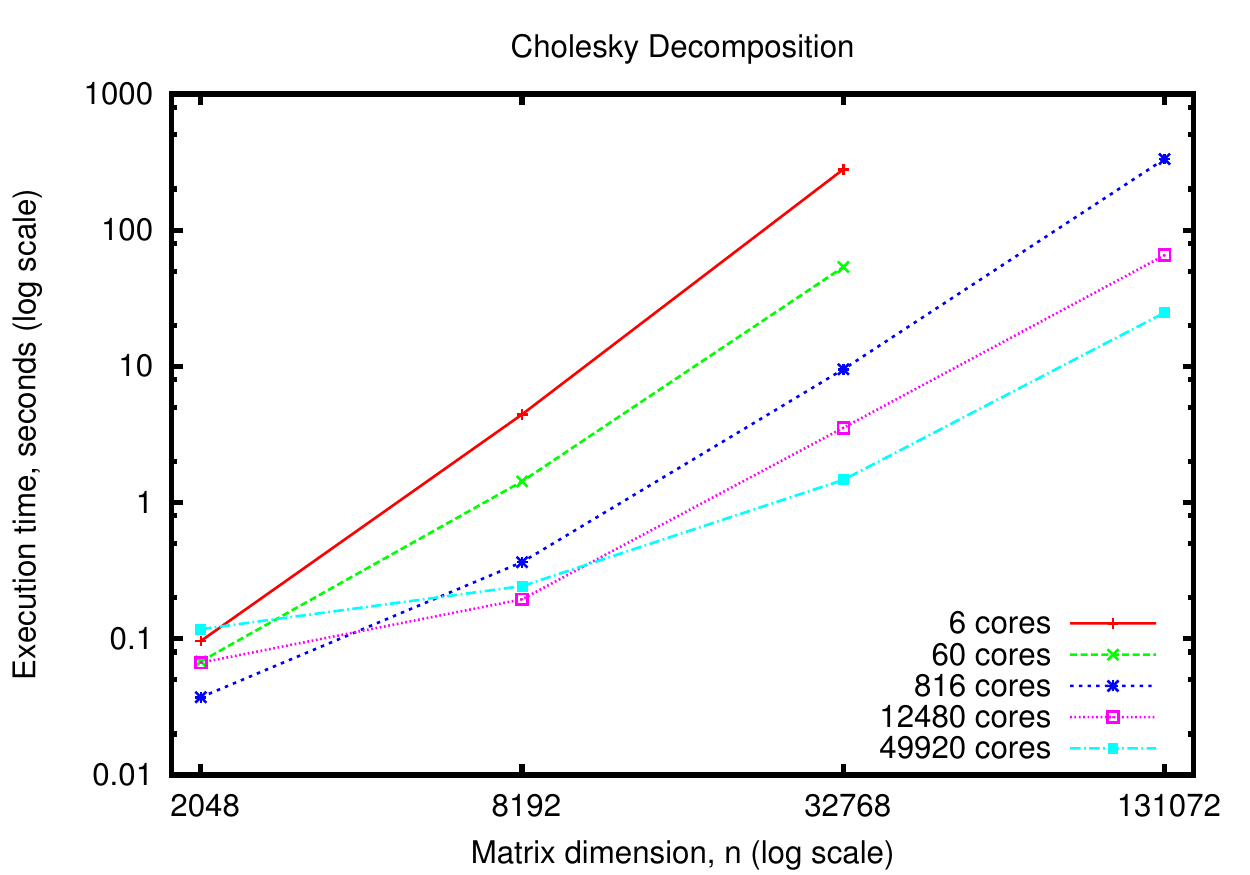}
  \caption{Runtimes as a function of $n$ for Cholesky decomposition on Hopper, for a variety of numbers of cores.  For 49920 cores, we used 24 cores per process; in the other cases we used 6 cores per process.}
  \label{fig:vary-n}
\end{figure}

As the matrix size $n$ increases, the arithmetic count of computations required for Cholesky decomposition increases as a function of $n^3$.  For small problem sizes, this increase is mitigated by the greater efficiency in computing with larger matrices.  Fig.~\ref{fig:vary-n} shows how runtime varies with $n$.

\subsection{Effect of number of cores per process}
\begin{figure}
  \centering
  \includegraphics[scale=1]{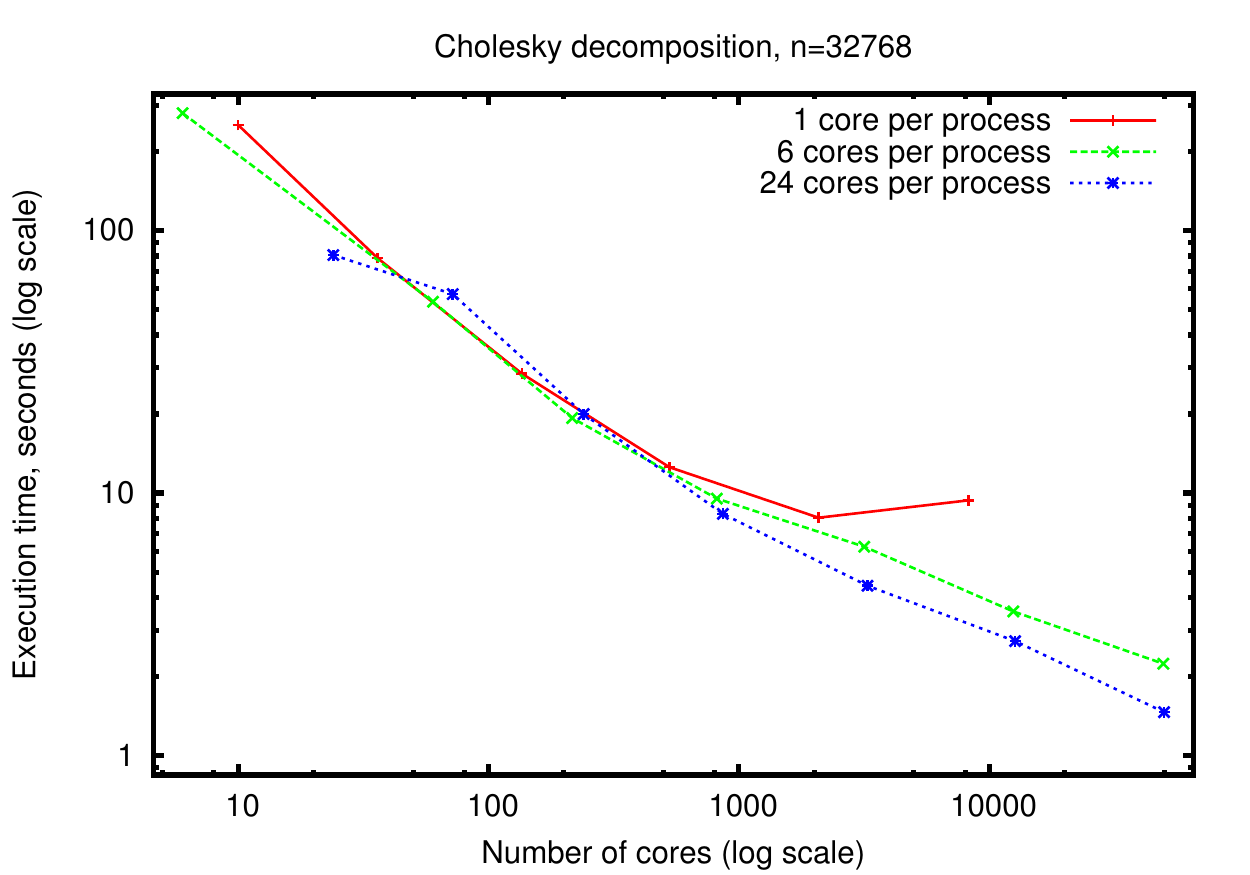}
  \caption{Runtimes for $32768\times32768$ Cholesky decomposition on Hopper using 1 core, 6 cores, or 24 cores (full node) per process.  For each data point, we use the optimal value of $h$.}
  \label{fig:vary-cpp}
\end{figure}

Our framework gives the user the freedom to choose how many cores to assign to each process, up to the number of cores on a node.  Whatever choice the user makes, all of the cores will be active most of the time.  When multiple cores are assigned to a single process, parallelism between cores comes from the threaded BLAS, whereas parallelism between the processes comes from our package.  Both use similar techniques to achieve parallelism.  The main difference is in the communication.  When running with many processes per node, each one is sending many small messages to processes on other nodes, but when running with one process per node one is more efficient in sending fewer, larger messages.  As Fig.~\ref{fig:vary-cpp} shows, the number of cores per process is not very important when using a small number of cores, up to about 480 cores, where the calculation is computation-bound.  As the number of cores increases, the calculation becomes communication-bound, and better performance is attained with fewer processes per node (more cores per process).  Note that the ideal choice of block size is affected by the number of cores per process, since efficiently using more cores requires larger blocks.  

\subsection{Using GPUs to speed up the linear algebra}
\begin{figure}
  \centering
  \includegraphics[scale=1]{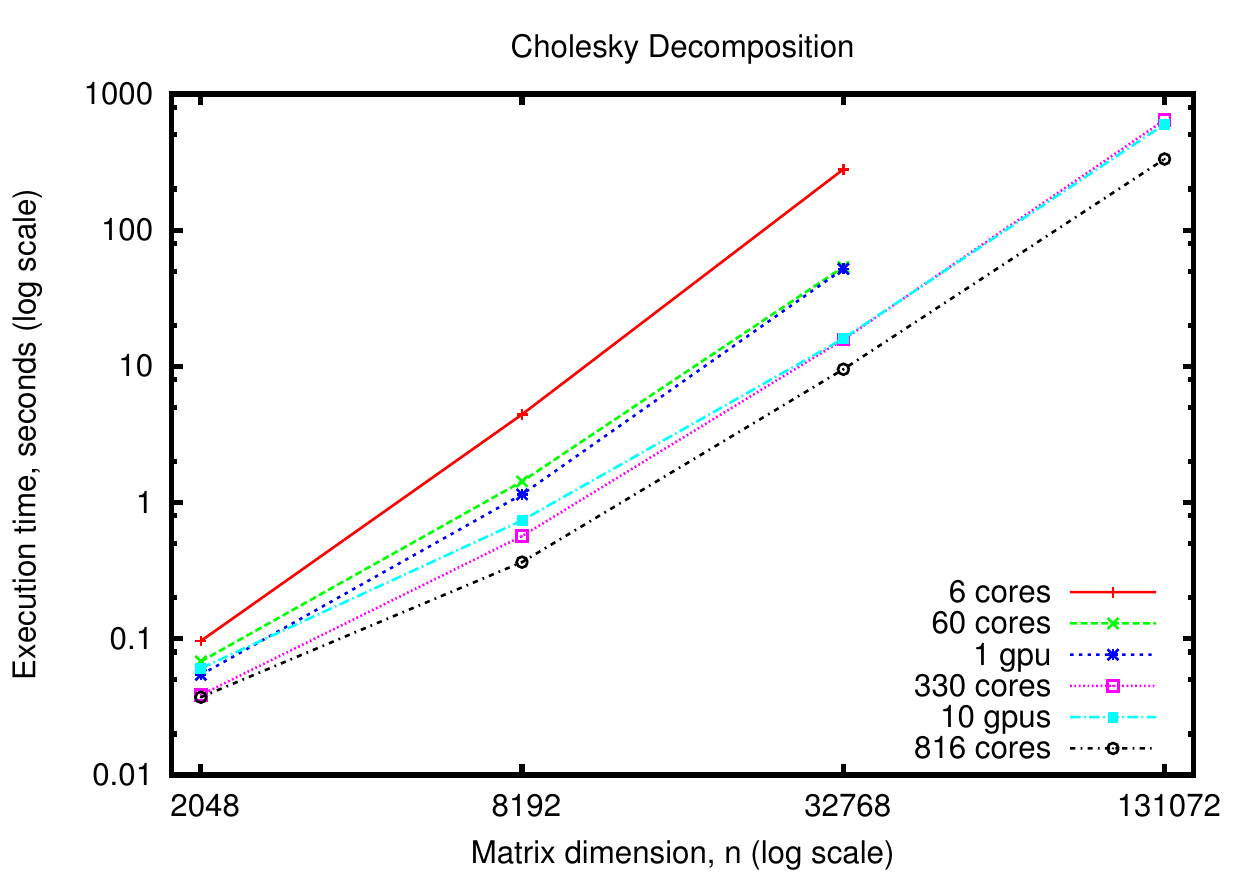}
  \caption{Runtimes as a function of $n$ for Cholesky decomposition using 1 and 10 GPUs on Dirac with results using Hopper for a variety of number of cores.  The CPU lines correspond to using 6 cores per process.}
  \label{fig:gpu}
\end{figure}

There is growing use of GPUs to speed up various computations, in particular linear algebra.  Our framework can be easily modified to run on a single GPU or a cluster of nodes with GPUs by using CUBLAS and MAGMA instead of BLAS and LAPACK.  We implemented Cholesky decomposition and tested on the NERSC machine Dirac, a small cluster with one NVIDIA Tesla C2050 GPU and 2 Intel 5530 CPUs per node.  Theoretically each GPU has the equivalent performance of about 60 cores of Hopper, although the interconnects are slower and so more problems are communication-bound.  Fig.~\ref{fig:gpu} compares the performance on 1 and 10 GPUs on Dirac to the performance on Hopper.  One GPU is roughly the same speed as 60 cores on Hopper (matching the theoretical result), whereas 10 GPUs gives roughly the same speed as 330 cores on Hopper (showing the slow-down due to communication).  When running on Dirac, the computation is entirely done on the GPUs; CPUs are only used for transferring data between nodes.  In principle one could try to divide the computation between the CPU and GPU, but, since the theoretical peak of the CPUs on each node is only 15\% that of the GPU, this would only yield slight performance improvements.

\section{Astrophysics example}
\label{sec:example}

\subsection{Background}

\begin{figure}
  \centering
  \includegraphics[scale=.5]{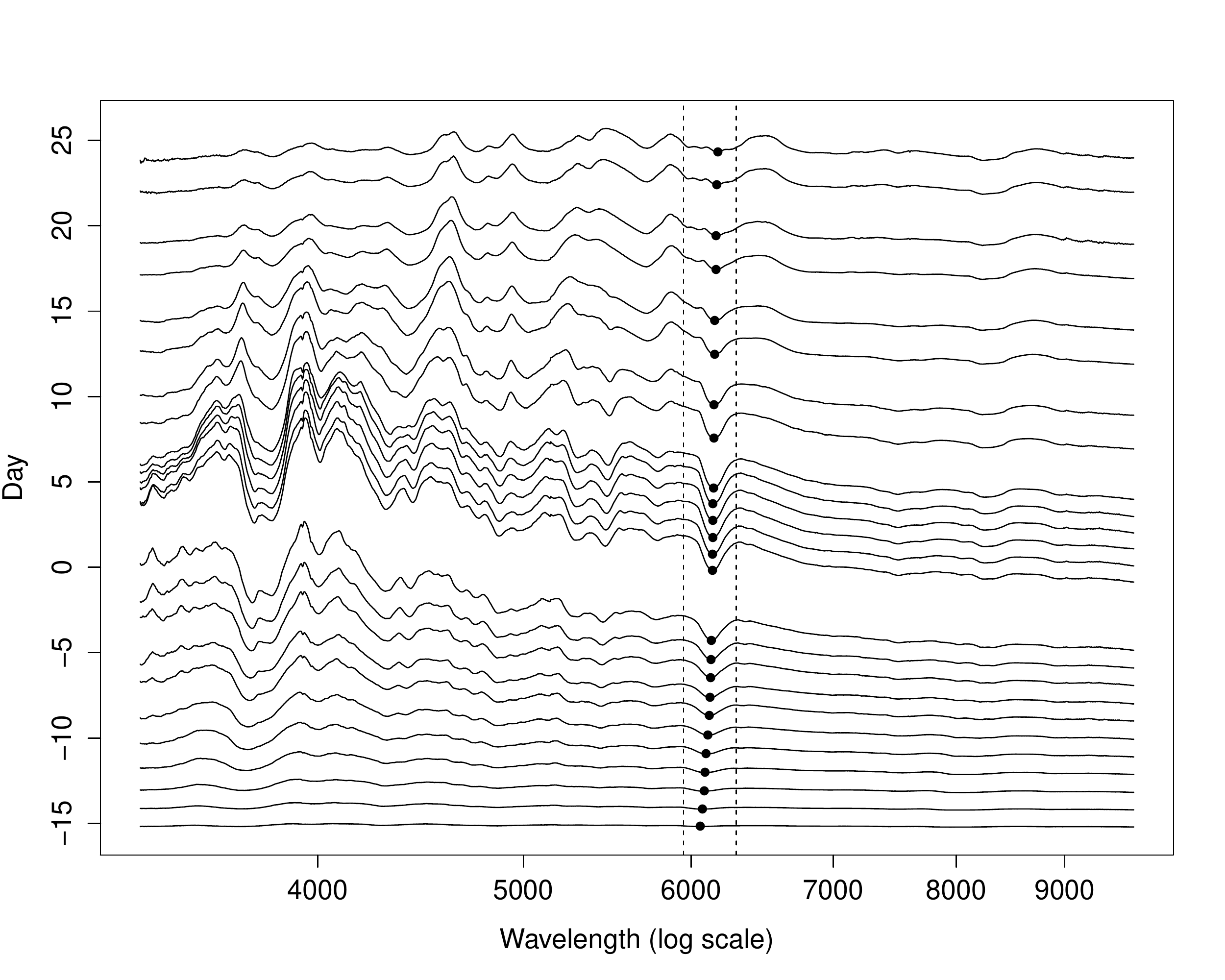}
  \caption{Flux observations for the Type~la supernova. Offset on the y-axis corresponds to time of observation in days. The scale for the flux measurements is not indicated due to the offset, but these measurements range from -0.003 to 1. Each spectrum contains measurements corresponding to 2691 wavelengths (\AA), shown on the x-axis. Wavelengths 5950 \AA~and 6300 \AA~are indicated by dotted vertical lines; this is the range over which we make predictions. The empirical minimum within this range for each spectrum is indicated with a solid dot.}
  \label{fig:data}
\end{figure}

Our example data set is the public spectrophotometric time series of the
Type~Ia supernova SN~2011fe \citep{pere:etal:2013}, obtained and reduced by
the Nearby Supernova Factory \citep{spie}.  The time series itself is a
sequence of spectra, each consisting of flux and flux error in units of
erg s$^{-1}$ cm$^{-2}$ \AA$^{-1}$ tabulated as a function of wavelength
in \AA.  Each spectrum was obtained on a different night.  There are 25 unique
spectra, each of which
contains 2691 flux (and flux error) measurements.  The total size of the
data set is thus 67,275 flux values.  The time coverage is not uniform, but
the wavelength grid is regularly spaced and the same from night to
night.  The flux values themselves are calibrated so that differences in
the brightness of the supernova from night to night and wavelength to
wavelength are physically meaningful. The data are shown in Fig.~\ref{fig:data}.

We are interested in obtaining a smoothed prediction of the flux of
SN~2011fe as a function of time and wavelength along with an estimate of
the prediction error.  The spectrum of a supernova contains broad
absorption and emission features whose appearance is the result of
physical processes and conditions in the expanding stellar ejecta.  The
widths, depths, and heights of such features change with time as the
supernova expands and cools.  The wavelengths of absorption feature
minima are examples of physically interesting quantities to extract from
spectral time series as a function of time.  These translate to a
characteristic ejecta velocity that provides an estimate of the kinetic
energy of the supernova explosion, something of great interest to those
that study exploding stars.

To demonstrate our algorithms and software on a real data set, we will
extract the velocity of the Si~II (singly ionized silicon) absorption
minimum typically found near 6150~\AA\ in Type~Ia supernovae.  This will
be done by finding the minimum absorption in the feature using the
smoothed representation of the supernova spectrum.  Realizations of the
spectrum, sampled from the posterior Gaussian process, will be used to
produce Monte Carlo error estimates on the position of the absorption
minimum.  Rather than measuring the position of each minimum only at
points where the data have been obtained (represented by the solid dots in Fig.~\ref{fig:data}), this procedure yields a smooth
estimate of the absorption minimum as a function of time
interpolated between observations, while also taking observation errors into account.

 
\subsection{Statistical model}

We model the flux measurements $Y_1, \ldots, Y_{67275}$ as being equal to a GP realization plus two error components: random effects for each phase (time point) and independent errors due to photon noise. We denote these three components by
$$Y_i = Z(t_i, w_i) + \alpha_{t_i} + \epsilon_{i},$$
where $t_i$ represents the time corresponding to $Y_i$ and $w_i$ the log wavelength, $\alpha_{t_i}$ is the random effect corresponding to time $t_i$, and $\epsilon_i$ is measurement error for the $i^{th}$ observation. The models for these components are
\begin{eqnarray*}
Z &\sim& GP(\mu(\cdot; \kappa, \lambda), \sigma^2 K(\cdot, \cdot; \rho_p, \rho_w)\\
\alpha_1, \ldots, \alpha_{25} &\stackrel{iid}{\sim}& N(0, \tau^2)\\
\epsilon_i &\sim& N(0, v_i), \quad \epsilon_1, \ldots, \epsilon_{67275} \mbox{ mutually independent}
\end{eqnarray*}

$Z$ has mean $\mu$, a function of time $t$ only, derived from a standard template Type Ia supernova spectral time series \citep{hsiao2007}, with $\kappa$ and $\lambda$ controlling scaling in magnitude and time. We take the correlation function to be a product of two Mat\'ern correlation functions, one for both the phase and log wavelength dimensions, each with smoothness parameter $\nu=2$.  Note that the flux error variances $v_i$ are known, leaving us with six parameters to be estimated.

\subsection{R code}
The first steps are to load the package, set up the inputs to the mean and covariance functions, and initialize the kriging problem object, called \texttt{prob}. Note that in this case the mean and covariance functions are provided by the package, but in general these would need to be provided by the user.
\begin{verbatim}
library(bigGP)

nProc <- 465
n <- nrow(SN2011fe)
m <- nrow(SN2011fe_newdata)
nu <- 2
inputs <- c(as.list(SN2011fe), as.list(SN2011fe_newdata), nu = nu)

prob <- krigeProblem$new("prob", numProcesses = nProc, h_n = NULL, 
   h_m = NULL, n = n, m = m, meanFunction = SN2011fe_meanfunc, 
   predMeanFunction = SN2011fe_predmeanfunc, 
   covFunction = SN2011fe_covfunc, 
   crossCovFunction = SN2011fe_crosscovfunc, 
   predCovFunction = SN2011fe_predcovfunc, inputs = inputs, 
   params = SN2011fe_initialParams, data = SN2011fe$flux, 
   packages = 'fields', parallelRNGpkg = "rlecuyer")
\end{verbatim}
We then maximize the log likelihood, followed by making the kriging predictions and generating a set of 1000 realizations from the conditional distribution of $Z$ given the observations and fixing the parameters at the maximum likelihood estimates. The predictions and realizations are over a grid, with days ranging from -15 to 24 in increments of 0.5 and wavelengths ranging from 5950 to 6300 in increments of 0.5. The number of prediction points is therefore $79 \times 701 = 55379$.
\begin{verbatim}
prob$optimizeLogDens(method = "L-BFGS-B", verbose = TRUE, 
    lower = rep(.Machine$double.eps, length(SN2011fe_initialParams)), 
    control = list(parscale = SN2011fe_initialParams))

pred <- prob$predict(ret = TRUE, se.fit = TRUE, verbose = TRUE)
realiz <- prob$simulateRealizations(r = 1000, post = TRUE, 
    verbose = TRUE)

\end{verbatim}

\subsection{Results}

\begin{figure}
  \centering
  \includegraphics[scale=.5]{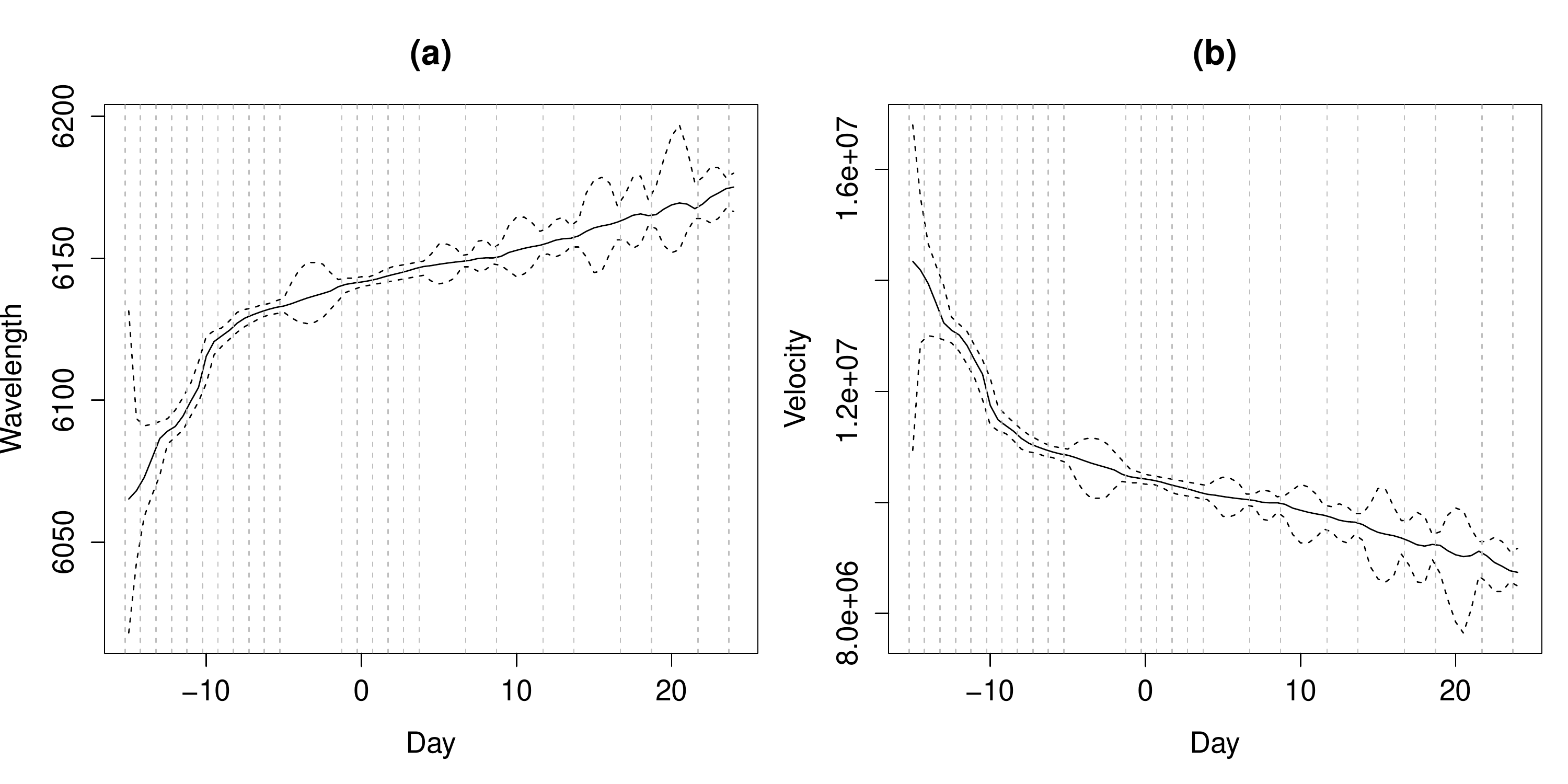}
  \caption{Posterior means (solid black lines) and pointwise 95\% posterior credible intervals (dotted black lines) for the wavelength (\AA) corresponding to the minimum flux (a) and the corresponding ejecta velocity (m/s) (b). The vertical gray dotted lines indicate observation times.}
  \label{fig:pred}
\end{figure}

The MLEs are $\hat\sigma^2 = 0.0071$, $\hat\rho_p = 2.33$, $\hat\rho_w = 0.0089$, $\hat\tau^2 = 2.6e-5$, and $\hat\kappa = 0.33$. Fig.~\ref{fig:pred} shows the posterior mean and pointwise 95\% posterior credible intervals for the wavelength corresponding to the minimum flux for each time point. These are calculated from the 1000 sampled posterior realizations of $Z$. For each realization, we calculate the minimizing wavelength for each time point. To translate each wavelength value to an ejecta velocity via the Doppler shift formula, we calculate $v = c(\lambda_R/w - 1)$ where $\lambda_R = 6355$ is the rest wavelength of an important silicon ion transition and $c$ is speed of light, $3 \times 10^8 m/s.$ The posterior mean and point wise 95\% posterior credible intervals for the ejecta velocities are shown in Fig.~\ref{fig:pred} (b). 

\section{Discussion}

Our software allows one to carry out standard Gaussian process calculations, such as likelihood maximization, prediction, and simulation of realizations, off the shelf, as illustrated in the astrophysics example, in situations in which calculations using threaded linear algebra on a single computer are not feasible because the calculations take too long or use too much memory.  The software enables a user to implement standard models and related models without approximations. One limitation of our implementation is that we do not do any pivoting, so Cholesky factorization of matrices that are not numerically positive definite fails. This occurred in the example when simulating realizations on fine grids of wavelength and phase. 


Of course with large datasets, one has the necessary statistical information to fit more complicated models, including hierarchical models, than the standard kriging methodology we implement. The package is designed to be extensible, providing a core set of distributed linear algebra functions common to Gaussian process calculations as an API usable from \texttt{R} without any knowledge of \texttt{C} or \texttt{MPI}. This allows others to implement other Gaussian process methodologies that rely on these functions. These might include Bayesian methods for nonstationary covariance models and spatio-temporal models among others. For example, MCMC updating steps might be done from the master process, with the linear algebra calculations done using the API. As can be seen from our timing results, a Cholesky decomposition for $32,768$ observations can be done in several seconds with a sufficient number of processors (a speed-up of 2-3 orders of magnitude relative to a single computational node), potentially enabling tens of thousands of MCMC updates. Or for a separable space-time model, the spatial and temporal covariance matrices could be manipulated separately using the package. One might also consider handling even larger datasets (e.g., millions of observations) by use of the core functions within the context of computationally-efficient methods such as low-rank approximations.

One useful extension of our approach would be to sparse matrices. Sparse matrices are at the core of computationally-efficient Markov random field spatial models \citep{Bane:etal:2003,Rue:etal:2009,Lind:etal:2011} and covariance tapering approaches \citep{Furr:etal:2006,Kauf:etal:2008,Sang:Huan:2012} and implementing a distributed sparse Cholesky decomposition could allow calculations for problems with hundreds of thousands or millions of locations.
To get a feel for what is possible, we benchmarked the PaStiX package \citep{Henon:2000:PPS:645612.662662} on Hopper on sparse matrices corresponding to a two-dimensional grid with five nonzeros per row. With one million locations, a Cholesky decomposition could be done in about 5 seconds using 96 cores (4 nodes) of Hopper; for 16 million locations, it took about 25 seconds using 384 cores (16 nodes).  Using more than this number of cores did not significantly reduce the running time.
We also note that with these improvements in speed for the Cholesky decomposition, computation of the covariance matrix itself may become the rate-limiting step in a spatial statistics calculation. Our current implementation takes a user-specified \texttt{R} function for constructing the covariance matrix and therefore does not exploit threading because \texttt{R} itself is not threaded. In this case, specifying more than one process per node would implicitly thread the covariance matrix construction, but additional work to enable threaded calculation of the covariance matrix may be worthwhile.

\section*{Acknowledgments}

This work is supported by the Director, Office of Science, Office of Advanced Scientific Computing Research, of the U.S. Department of Energy under Contract No. AC02-05CH11231. This research used resources of the National Energy Research Scientific Computing Center.

\bibliographystyle{plainnat}
\bibliography{draftArxiv}

\end{document}